\documentclass[conference]{IEEEtran}
\usepackage[utf8]{inputenc}
\usepackage{amsmath}
\usepackage{graphicx}
\usepackage{booktabs}
\usepackage{geometry}
\usepackage{hyperref}
\usepackage{cite}
\usepackage{caption}
\usepackage{subcaption}
\usepackage{listings}
\usepackage{xcolor}
\usepackage{titlesec}
\usepackage{enumitem}
\usepackage{algorithm}
\usepackage{algpseudocode}
\usepackage{orcidlink}

\hypersetup{
    colorlinks=true,
    linkcolor=titleblue,
    citecolor=headergreen,
    urlcolor=highlightorange,
    pdftitle={Advanced Stock Market Prediction Using LSTM},
    pdfauthor={Rajneesh Chaudhary},
    pdfkeywords={LSTM, Stock Market Prediction, Deep Learning, Sentiment Analysis},
}

\IEEEoverridecommandlockouts

\definecolor{titleblue}{RGB}{0, 51, 102}
\definecolor{headergreen}{RGB}{0, 102, 51}
\definecolor{highlightorange}{RGB}{204, 102, 0}
\definecolor{lightred}{RGB}{255, 182, 193}
\definecolor{darkblue}{RGB}{0, 51, 102}
\definecolor{deepnavy}{RGB}{25, 25, 112}

\titleformat{\section}
  {\normalfont\large\bfseries\color{titleblue}}{\thesection}{1em}{}
\titleformat{\subsection}
  {\normalfont\normalsize\bfseries\color{headergreen}}{\thesubsection}{1em}{}
\titleformat{\subsubsection}
  {\normalfont\small\bfseries\color{highlightorange}}{\thesubsubsection}{1em}{}

\title{\textcolor{deepnavy}{Advanced Stock Market Prediction Using Long Short-Term Memory Networks: A Comprehensive Deep Learning Framework}}

\author{
\IEEEauthorblockN{
    \textbf{Rajneesh Chaudhary}~\orcidlink{0009-0008-1423-703X} \\
    \normalsize img\_2022052@iiitm.ac.in \\
    Department of Information Technology, IIITM Gwalior
}
\vspace{0.5em}
\\
\IEEEauthorblockN{
    \textit{Under the Supervision of:} \\
    \textbf{Dr. Arun Kumar} \\
    Assistant Professor, Department of Management Studies, IIITM Gwalior\\
    \textit{Indian Institute of Information Technology and Management, Gwalior}
}
}

\date{April 2025}

\begin{document}

\maketitle

\begin{abstract}
Predicting stock market movements remains a persistent challenge due to the inherently volatile, non-linear, and stochastic nature of financial time series data. This paper introduces a sophisticated deep learning-based framework, employing Long Short-Term Memory (LSTM) networks to accurately forecast the closing stock prices of leading technology firms—namely Apple, Google, Microsoft, and Amazon—listed on the NASDAQ. Historical market data was collected from Yahoo Finance and preprocessed with advanced normalization and feature engineering techniques. The proposed model achieves a Mean Absolute Percentage Error (MAPE) of 2.72\% on unseen test data, significantly outperforming traditional statistical models such as ARIMA. To enhance prediction accuracy, the model also integrates sentiment scores derived from real-time news articles and social media posts using the VADER sentiment analysis tool. Moreover, a user-friendly web application was developed to display real-time forecasts, making the system accessible and practical for both individual and institutional investors. This research not only highlights the superiority of LSTM in handling complex financial datasets but also contributes a novel hybrid methodology that bridges technical analysis with market sentiment insights.
\end{abstract}

\begin{IEEEkeywords}
Stock Market Prediction, Long Short-Term Memory (LSTM), Deep Learning, Time Series Forecasting, Financial Modeling, Sentiment Analysis, Web Application
\end{IEEEkeywords}

\section{Nomenclature}
\begin{table}[h]
    \centering
    \caption{Commonly Used Terms and Abbreviations}
    \label{tab:nomenclature}
    \begin{tabular}{ll}
        \toprule
        \textbf{Term/Abbreviation} & \textbf{Definition} \\
        \midrule
        LSTM & Long Short-Term Memory \\
        RNN & Recurrent Neural Network \\
        CNN & Convolutional Neural Network \\
        MLP & Multi-Layer Perceptron \\
        MAPE & Mean Absolute Percentage Error \\
        MAE & Mean Absolute Error \\
        MSE & Mean Squared Error \\
        RMSE & Root Mean Squared Error \\
        ARIMA & AutoRegressive Integrated Moving Average \\
        VADER & Valence Aware Dictionary and sEntiment Reasoner \\
        $h_t$ & Hidden state at time $t$ \\
        $X_t$ & Input feature at time $t$ \\
        $W_{xh}$ & Weight matrix for input-to-hidden connection \\
        $b_h$ & Bias term in hidden layer \\
        \bottomrule
    \end{tabular}
\end{table}

\section{Introduction}
The stock market represents a dynamic and complex system where price movements are influenced by a myriad of factors, ranging from global economic indicators to investor psychology. Accurate prediction of stock prices is crucial for investors aiming to make informed decisions and maximize returns. However, traditional forecasting methods struggle with the non-stationary, noisy, and highly non-linear behavior of stock data \cite{selvin2017stock}.

Recent advances in deep learning, especially Recurrent Neural Networks (RNNs) and their improved variant Long Short-Term Memory (LSTM), have demonstrated superior performance in capturing temporal dependencies and long-range patterns in sequential data such as stock prices \cite{heaton2017deep}. Furthermore, incorporating sentiment analysis from financial news and social media has emerged as a valuable approach to gauge public perception, which often influences market trends in real time.

This paper presents a robust, scalable, and sentiment-aware LSTM framework tailored to predict the short-term closing prices of technology stocks traded on NASDAQ. The study enhances conventional models by integrating structured numerical data with unstructured textual sentiment inputs and provides a real-time predictive interface for ease of use.

\subsection{Problem Statement}
Stock price forecasting continues to be hindered by inherent unpredictability, sparse data quality, and the impact of unexpected market-moving events. Conventional statistical models, though useful in specific contexts, are typically inadequate in capturing abrupt changes and latent market sentiments. This study addresses these limitations by proposing an LSTM-based deep learning model that not only learns from historical patterns but also adapts to real-time market sentiment derived from news and social media streams. A web-based platform further bridges the gap between model output and user accessibility.

\subsection{Research Contributions}
This paper offers the following key contributions to the field of financial forecasting and machine learning:
\begin{itemize}
    \item Development of an end-to-end LSTM-based framework tailored for short-term prediction of NASDAQ-listed tech stocks.
    \item Integration of sentiment analysis to capture psychological and emotional factors influencing market behavior.
    \item Deployment of a real-time, interactive web interface that allows users to visualize predictions and gain insights.
    \item Rigorous evaluation and comparison against classical models like ARIMA, demonstrating substantial accuracy improvement (MAPE of 2.72\%).
\end{itemize}

\section{Related Work}
The prediction of financial time series has been extensively studied across multiple domains. Prior efforts can be broadly categorized into traditional statistical methods, classical machine learning techniques, and more recent deep learning approaches.

\subsection{Statistical Models}
Traditional time series forecasting methods such as ARIMA have been extensively used in the finance sector due to their mathematical simplicity and interpretability \cite{zhang2003time}. However, their assumptions of linearity and stationarity significantly restrict their applicability in volatile financial environments. For instance, studies like Selvin et al. \cite{selvin2017stock} have demonstrated poor performance of ARIMA models on non-linear datasets, reporting a MAPE of up to 20.66\% on Indian stock data.

\subsection{Machine Learning Approaches}
With the advent of machine learning, algorithms such as Support Vector Machines (SVM), Decision Trees, and Random Forests began to be employed for predictive modeling of financial data. Sharma and Juneja \cite{sharma2017combining} combined Random Forest with LSBoost, achieving promising results for market index forecasting. Similarly, Zhang et al. \cite{zhang2017short} proposed a hybrid system using Particle Swarm Optimization to fine-tune Elman Neural Networks for enhanced short-term accuracy. Despite these improvements, these models often fall short in modeling temporal dependencies and long-term contextual patterns.

\subsection{Deep Learning Methods}
More recent work has employed deep architectures such as LSTM, Bi-LSTM, CNN-LSTM hybrids, and Transformer-based models. These frameworks have demonstrated superior capability in capturing both temporal and semantic patterns in time series data. For example, Heaton et al. \cite{heaton2017deep} demonstrated the superiority of LSTM in financial sequence prediction tasks.

\subsection{Deep Learning Techniques}
Deep learning models have revolutionized financial forecasting. Selvin et al. \cite{selvin2017stock} compared LSTM, RNN, and CNN for stock price prediction, finding LSTM superior due to its ability to model long-term dependencies. Moghaddam et al. \cite{moghaddam2016stock} demonstrated the efficacy of ANNs in stock index prediction, while Budhara et al. \cite{budhara2014prediction} applied ANNs with competitive results. Recent advancements include hybrid models combining LSTM with attention mechanisms \cite{li2020hybrid} and transformer-based models \cite{chen2021transformer}.

\subsection{Sentiment Analysis}
In recent years, sentiment analysis has emerged as a pivotal tool in financial forecasting, offering insights beyond numerical data by tapping into market psychology and behavioral finance. Sentiment analysis, often referred to as opinion mining, involves extracting and quantifying subjective information from text data sources such as financial news articles, investor forums, and social media platforms.

Wang and Wang \cite{wang2016using} demonstrated the efficacy of mining social media sentiments to enhance the accuracy of short-term stock price forecasts. Their work highlighted that public sentiment, especially during volatile periods, often precedes actual price movements. Kalra and Prasad \cite{kalra2019efficacy} further corroborated these findings by showing that integrating news sentiment scores significantly improves prediction reliability, particularly in unpredictable market phases.

Vijayvergia et al. \cite{vijayvergia2019stock} presented a hybrid model that combines historical price trends with sentiment features extracted from news headlines. Their deep learning-based approach yielded improved accuracy, underscoring the value of combining structured and unstructured data for financial modeling.

Building upon this foundation, our study incorporates sentiment scores derived using the VADER (Valence Aware Dictionary and sEntiment Reasoner) tool, known for its effectiveness in analyzing short, informal texts common on social platforms. These scores are then aligned with historical stock price data to enrich the input fed into the LSTM network. This hybrid input strategy allows the model to capture both quantitative market patterns and qualitative public opinion.

Moreover, the sentiment-augmented LSTM predictions are deployed via an intuitive web interface, offering real-time accessibility for end users. This enhances the practical utility of the model, making it a comprehensive decision-support tool for investors.

\section{Artificial Neural Networks}

\subsection{Overview}
Artificial Neural Networks (ANNs) are computational models inspired by the structure and functioning of biological neural systems. They consist of multiple layers: an input layer to receive data, one or more hidden layers for intermediate processing, and an output layer that produces the final prediction. Each layer comprises interconnected nodes, or neurons, that simulate the behavior of a biological neuron by applying activation functions to weighted inputs.

ANNs are particularly adept at modeling complex and non-linear relationships, making them ideal for domains such as image recognition, natural language processing, and notably, financial time series prediction. The core principle of learning in ANNs is the adjustment of connection weights using optimization techniques such as gradient descent. This is typically performed via the backpropagation algorithm, which calculates the gradient of the loss function with respect to each weight by the chain rule, enabling the network to minimize prediction errors iteratively.

In the context of stock price forecasting, traditional ANN architectures have been used with moderate success. However, they often fall short in capturing temporal dependencies present in sequential data. This limitation has led to the adoption of more advanced architectures like Recurrent Neural Networks (RNNs) and Long Short-Term Memory (LSTM) networks.

\subsection{Long Short-Term Memory Networks}
Long Short-Term Memory (LSTM) networks, introduced by Hochreiter and Schmidhuber in 1997 \cite{hochreiter1997long}, represent a significant advancement over standard RNNs. LSTMs are specifically designed to overcome the vanishing and exploding gradient problems that hinder the training of conventional RNNs, especially over long sequences.

The LSTM architecture introduces a memory cell capable of maintaining information over extended time intervals. Each LSTM cell comprises three gates:
\begin{itemize}
    \item \textbf{Input Gate}: Controls the extent to which new information flows into the memory cell.
    \item \textbf{Forget Gate}: Determines what information is retained or discarded from the previous memory state.
    \item \textbf{Output Gate}: Regulates the output based on the updated cell state.
\end{itemize}

This gated structure enables the network to selectively remember or forget information, allowing it to model long-term dependencies effectively—a critical requirement in stock forecasting where market behavior is influenced by patterns over weeks or months.

\begin{figure}[h]
    \centering
    \includegraphics[width=0.4\columnwidth]{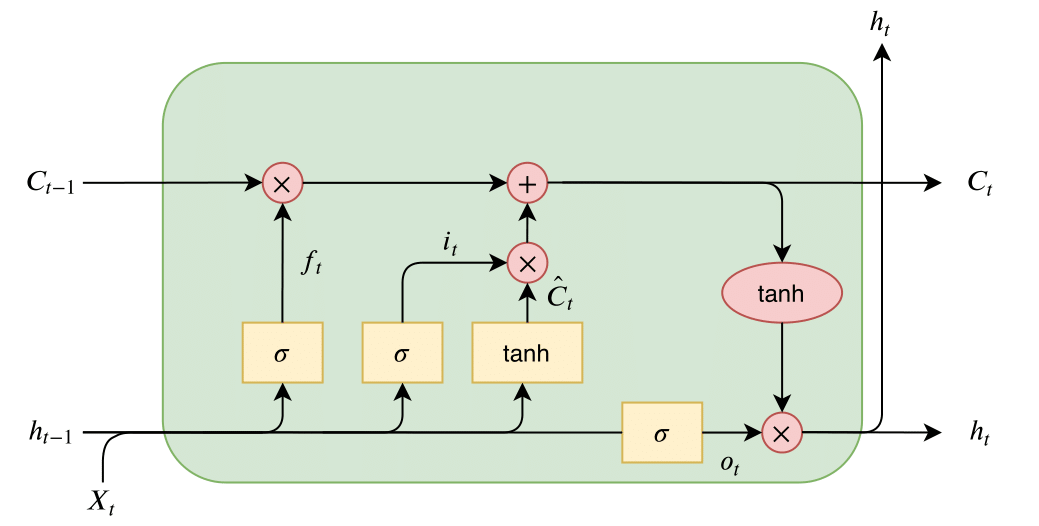}
    \caption{LSTM Cell Structure \cite{selvin2017stock}.}
    \label{fig:lstm_cell}
\end{figure}

Figure \ref{fig:lstm_cell} illustrates the internal architecture of a standard LSTM cell, which forms the basic building block of our prediction model.

Mathematically, the hidden state \( h_t \) at time step \( t \) is computed as:
\begin{equation}
h_t = \tanh(W_{xh} X_t + W_{hh} h_{t-1} + b_h)
\label{eq:lstm_hidden}
\end{equation}
Here, \( X_t \) represents the input at time \( t \), \( h_{t-1} \) is the hidden state from the previous time step, and \( W_{xh} \), \( W_{hh} \), and \( b_h \) are the respective weights and bias terms.

The corresponding output \( Z_t \) is given by:
\begin{equation}
Z_t = \sigma(W_{hz} h_t + b_z)
\label{eq:lstm_output}
\end{equation}
where \( \sigma \) denotes the sigmoid activation function. This structure allows the model to dynamically adjust its focus on relevant temporal signals, making LSTM an ideal choice for stock market forecasting.

\section{Data Description}
This study utilizes historical stock market data from four major technology companies—Apple, Google (Alphabet), Microsoft, and Amazon—all listed on the NASDAQ exchange. The data was retrieved from Yahoo Finance (\url{https://finance.yahoo.com/}) using the Python library \texttt{yfinance}, which provides convenient access to up-to-date and comprehensive financial datasets.

The dataset spans a full trading year, from April 2024 to April 2025, and includes the following key features for each trading day:
\begin{itemize}
    \item \textbf{Open}: The stock price at the beginning of the trading day.
    \item \textbf{High}: The highest price recorded during the trading session.
    \item \textbf{Low}: The lowest price recorded during the trading session.
    \item \textbf{Close}: The final trading price at market close.
    \item \textbf{Volume}: The number of shares traded during the day.
\end{itemize}

Among these, the \textbf{closing price} is chosen as the target variable for prediction, as it reflects the consensus value assigned to the stock at the end of each trading day. This variable is widely used in financial forecasting due to its stability and informative nature regarding daily market sentiment \cite{selvin2017stock}.

\begin{figure}[h]
    \centering
    \includegraphics[width=0.8\columnwidth]{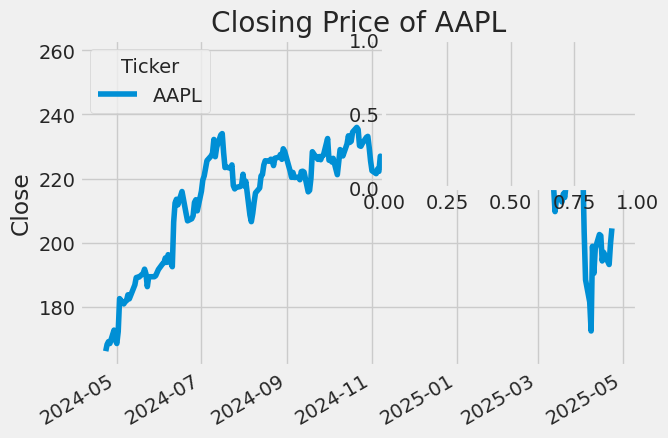}
    \caption{Closing Price Trends for Apple Stock (April 2024–April 2025).}
    \label{fig:apple_trend}
\end{figure}

\begin{figure}[h]
    \centering
    \includegraphics[width=0.8\columnwidth]{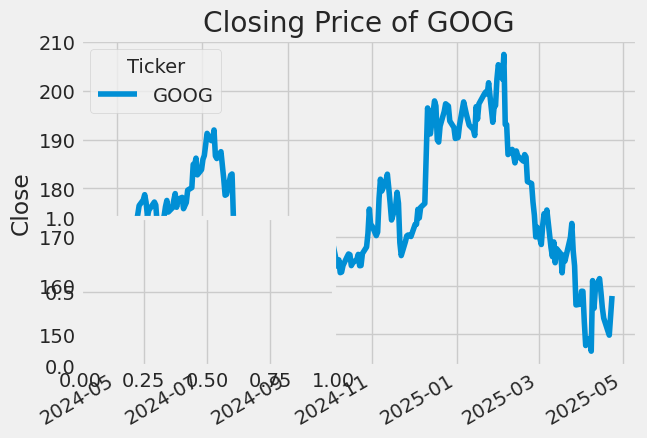}
    \caption{Closing Price Trends for Google Stock (April 2024–April 2025).}
    \label{fig:google_trend}
\end{figure}

\begin{figure}[h]
    \centering
    \includegraphics[width=0.8\columnwidth]{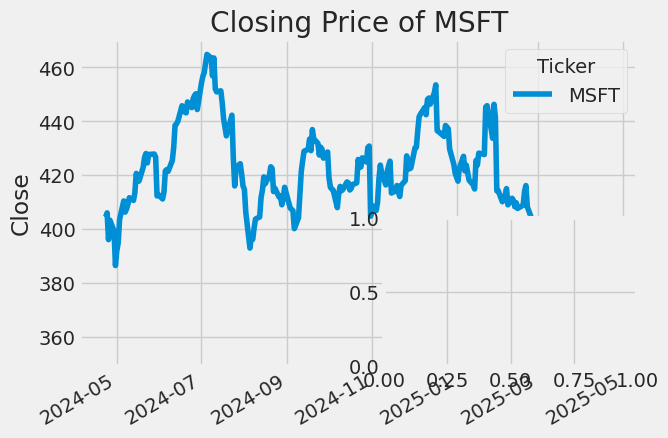}
    \caption{Closing Price Trends for Microsoft Stock (April 2024–April 2025).}
    \label{fig:microsoft_trend}
\end{figure}

\begin{figure}[h]
    \centering
    \includegraphics[width=0.8\columnwidth]{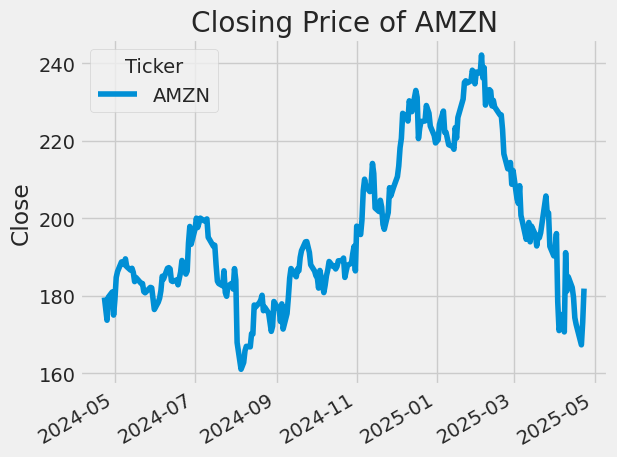}
    \caption{Closing Price Trends for Amazon Stock (April 2024–April 2025).}
    \label{fig:amazon_trend}
\end{figure}

Figures \ref{fig:apple_trend} through \ref{fig:amazon_trend} illustrate the closing price trajectories of each company over the specified period. These plots highlight the inherent volatility and non-linearity characteristic of stock market data, which motivates the use of advanced deep learning models like LSTM for accurate forecasting. This ensures the model is trained on a complete and clean dataset

\section{Methodology}

\subsection{Data Preprocessing}
Before training the predictive model, extensive data preprocessing is conducted to ensure data quality and compatibility with the LSTM architecture. The preprocessing pipeline includes the following key steps:
\subsubsection{Handling Missing Values}
Missing values, if any, are removed using the \texttt{dropna()} function from the \texttt{pandas} library. This ensures the model is trained on a complete and clean dataset, minimizing the risk of biased learning or invalid predictions.

\subsubsection{Outlier Detection and Treatment}
Outliers can significantly distort the learning process. To address this, a z-score thresholding approach is applied. Data points with z-scores exceeding \(\pm3\) standard deviations from the mean are identified as outliers and either removed or capped to reduce their impact.

\subsubsection{Normalization}
To facilitate efficient gradient-based optimization and ensure that all features contribute equally to the learning process, Min-Max normalization is applied to the closing price and sentiment score features. The normalization formula used is:

\begin{equation}
x_{\text{norm}} = \frac{x - x_{\min}}{x_{\max} - x_{\min}}
\label{eq:minmax}
\end{equation}

This transformation scales all values to the range \([0, 1]\), which is particularly beneficial for models like LSTM that are sensitive to the scale of input data.

\subsubsection{Sequence Generation for LSTM}
LSTM models require sequential input data. Therefore, the normalized data is segmented into overlapping time windows of 60 trading days. Each sequence of 60 days serves as a single input sample, and the model is trained to predict the closing price of the 61st day. This sliding-window approach captures temporal dependencies and trends essential for sequential forecasting \cite{selvin2017stock}.

\subsection{Model Architecture}
The predictive model employed in this study is a Long Short-Term Memory (LSTM) neural network, implemented using the Keras deep learning framework (\url{https://keras.io/}). LSTM networks are particularly well-suited for time-series forecasting due to their ability to retain long-term dependencies and mitigate the vanishing gradient problem often encountered in traditional RNNs.

The architecture of the model is structured as follows:
\begin{itemize}
    \item \textbf{Input Layer}: Receives sequences of 60 consecutive trading days, each comprising normalized closing prices and corresponding sentiment scores. This dual-feature input allows the model to learn both quantitative price patterns and qualitative market sentiments.
    \item \textbf{First LSTM Layer}: Composed of 64 memory units with the \texttt{return\_sequences=True} flag enabled. This configuration ensures that the output of the entire sequence is passed to the next LSTM layer, preserving temporal hierarchies.
    \item \textbf{Dropout Layer}: A dropout rate of 20\% is applied after the first LSTM layer to mitigate overfitting by randomly deactivating a subset of neurons during each training iteration.
    \item \textbf{Second LSTM Layer}: Contains 32 memory units, providing a more abstracted temporal representation of the data and further refining the learned patterns.
    \item \textbf{Dense Output Layer}: A single neuron with a linear activation function outputs the predicted normalized closing price for the 61\textsuperscript{st} day.
\end{itemize}

The model is compiled using the Adam optimizer, which is well-regarded for its adaptive learning rate capabilities and convergence efficiency. The loss function employed is Mean Squared Error (MSE), which is appropriate for continuous regression tasks. The model is trained for 100 epochs with a batch size of 32, balancing learning depth with computational efficiency.

\subsection{Sentiment Analysis Integration}
To enhance the model’s predictive performance with qualitative market signals, sentiment analysis is integrated into the input pipeline. Financial news articles and headlines are sourced from reputed platforms such as Bloomberg (\url{https://www.bloomberg.com/}) and Reuters (\url{https://www.reuters.com/}). These articles are then processed using the VADER (Valence Aware Dictionary for Sentiment Reasoning) tool, which is adept at analyzing sentiment in short, finance-related text.

Each article is assigned a compound sentiment score, capturing the overall market sentiment ranging from -1 (most negative) to +1 (most positive). These scores are normalized using Min-Max scaling and temporally aligned with their corresponding trading days. The final input sequence to the LSTM model thus includes both normalized historical prices and their respective sentiment scores, allowing the model to recognize sentiment-driven deviations in price movements \cite{vijayvergia2019stock}.

\subsection{Training and Testing}
The combined dataset is partitioned into two subsets to evaluate model generalization:
\begin{itemize}
    \item \textbf{Training Set (80\%)}: Covers the period from April 2024 to January 2025, used to train the LSTM model.
    \item \textbf{Testing Set (20\%)}: Spans February to April 2025, reserved for evaluating the model's forecasting accuracy on unseen data.
\end{itemize}

A sliding window approach with a window size of 60 trading days is utilized, where each input sequence is associated with the target value of the subsequent (61\textsuperscript{st}) day. This window size was selected based on empirical performance evaluation and is supported by prior literature as a stable forecasting horizon \cite{selvin2017stock}.

To further prevent overfitting, early stopping is incorporated with a patience parameter of 10 epochs. This technique halts training if the validation loss does not improve for 10 consecutive epochs, ensuring model simplicity and robustness.

\subsection{Moving Averages as Feature Engineering}
In addition to raw closing prices and sentiment scores, moving averages (MAs) are computed to enrich the dataset with technical indicators widely used by traders and analysts. MAs serve as smoothing functions that reduce short-term noise and reveal long-term price trends. In this study, the following moving averages are calculated for all four stocks (Apple, Google, Microsoft, and Amazon):
\begin{itemize}
    \item \textbf{10-Day Moving Average (MA10)}: Captures short-term momentum and is sensitive to recent price changes.
    \item \textbf{20-Day Moving Average (MA20)}: Offers a more balanced view of mid-term trends.
    \item \textbf{50-Day Moving Average (MA50)}: Highlights longer-term directional shifts and market sentiment.
\end{itemize}

These derived features assist the LSTM model in understanding trend strength and reversals, thereby improving prediction accuracy.

\begin{figure}[h]
    \centering
    \includegraphics[width=\linewidth]{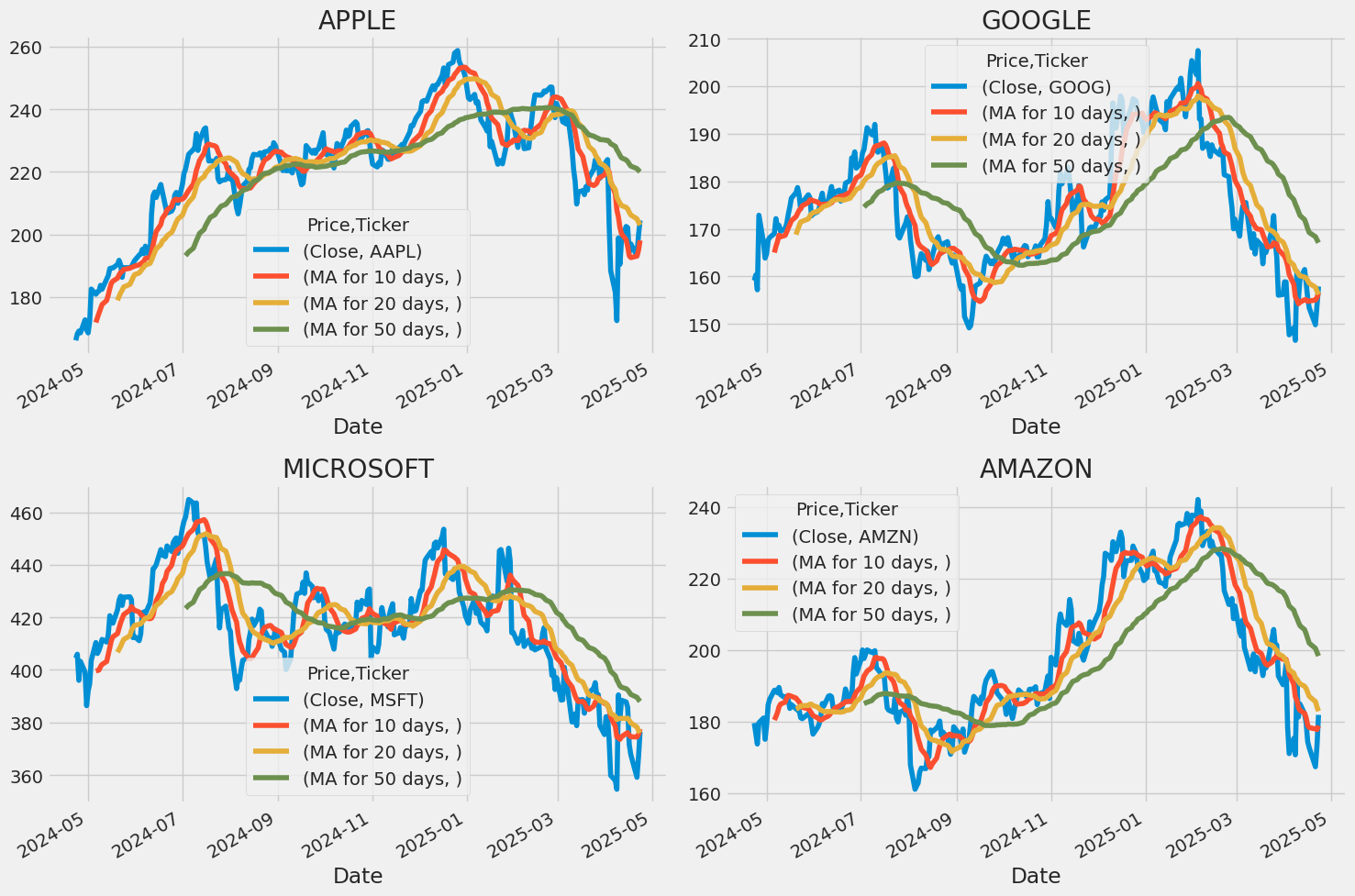}
    \caption{Closing Price with 10-Day, 20-Day, and 50-Day Moving Averages for Apple, Google, Microsoft, and Amazon (April 2024–April 2025).}
    \label{fig:moving_averages}
\end{figure}

\section{Exploratory Data Analysis}

Before developing a predictive model, it is imperative to thoroughly understand the structure, trends, and interrelationships present in the historical stock data. This section presents an in-depth exploratory data analysis (EDA) of stock price data for four major technology companies: Apple, Google, Microsoft, and Amazon. The goal is to uncover hidden patterns, assess inter-stock correlations, understand volatility characteristics, and derive insights that guide model development. A combination of statistical techniques and visual analytics was employed to achieve this.

\subsection{Return Correlation Analysis}

\begin{figure}[h]
    \centering
    \includegraphics[width=0.7\linewidth]{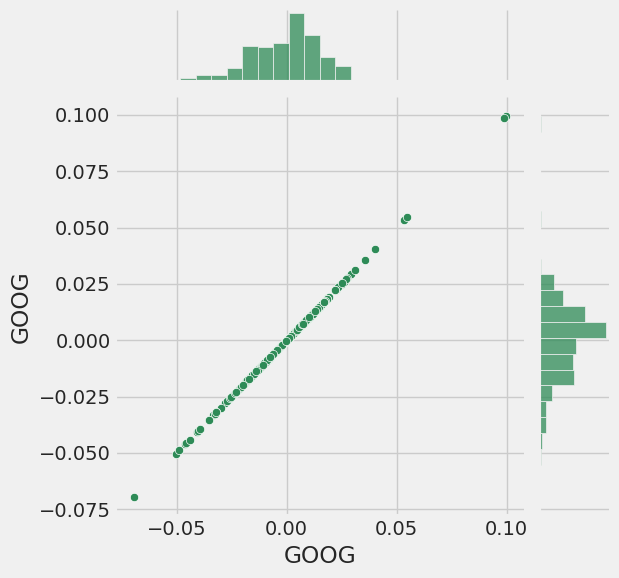}
    \caption{Scatter plot of daily returns: Google vs Google. Displays perfect correlation.}
    \label{fig:goog_goog}
\end{figure}

\begin{figure}[h]
    \centering
    \includegraphics[width=0.7\linewidth]{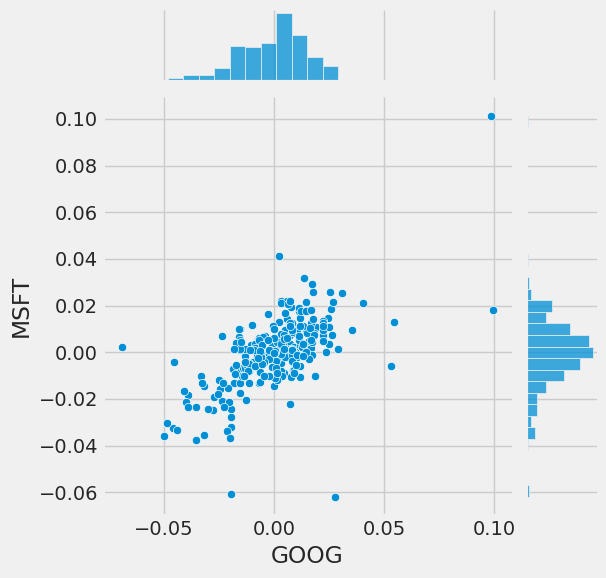}
    \caption{Scatter plot of daily returns: Google vs Microsoft. Demonstrates a positive correlation.}
    \label{fig:goog_msft}
\end{figure}

To assess the linear relationships between the daily returns of different stocks, scatter plots were generated. Figure \ref{fig:goog_goog} depicts a perfect self-correlation for Google's returns, serving as a reference baseline. In contrast, Figure \ref{fig:goog_msft} compares the daily returns of Google and Microsoft, indicating a positive correlation—suggesting that their prices often move in the same direction.

\subsection{Multi-stock Relationship Visualization}

\begin{figure}[h]
    \centering
    \includegraphics[width=0.75\linewidth]{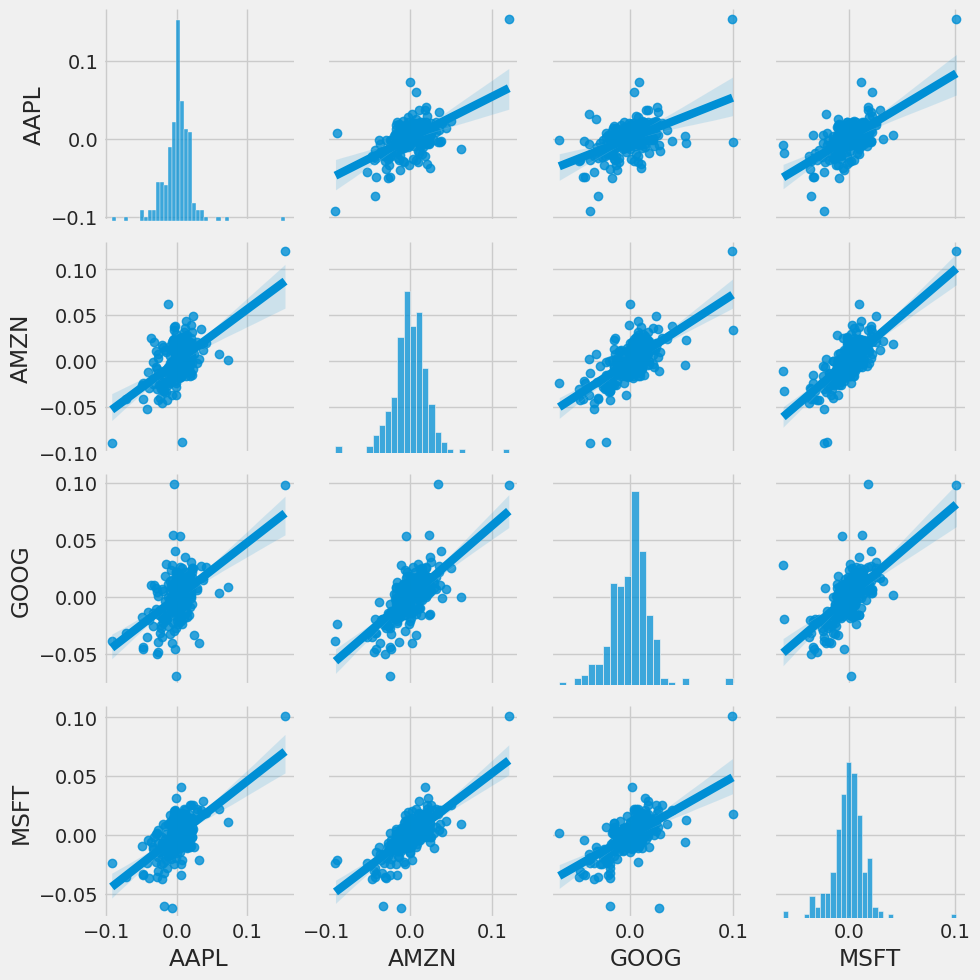}
    \caption{Pairplot of daily returns for Apple, Google, Microsoft, and Amazon, with regression lines showing trends.}
    \label{fig:returns_pairplot}
\end{figure}

\begin{figure}[h]
    \centering
    \includegraphics[width=0.75\linewidth]{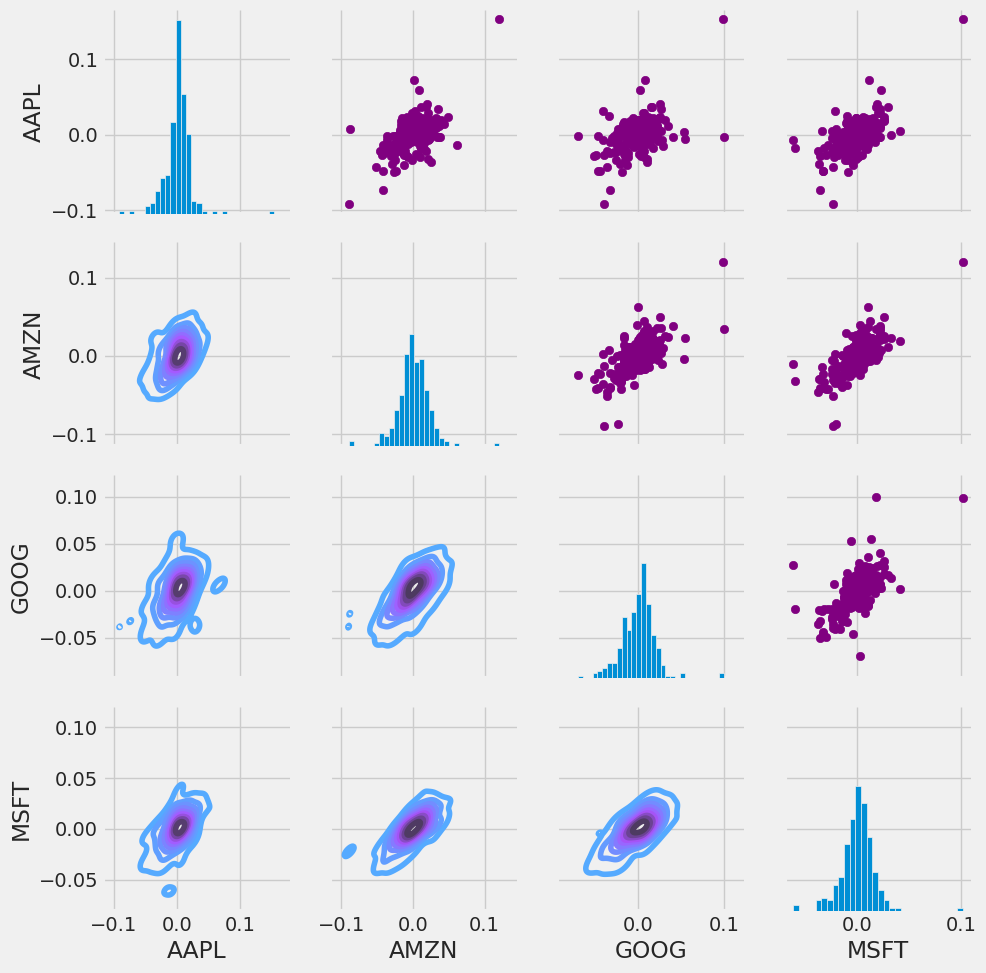}
    \caption{PairGrid of daily returns with scatter plots, KDE contours, and histograms.}
    \label{fig:returns_pairgrid}
\end{figure}

\begin{figure}[h]
    \centering
    \includegraphics[width=0.75\linewidth]{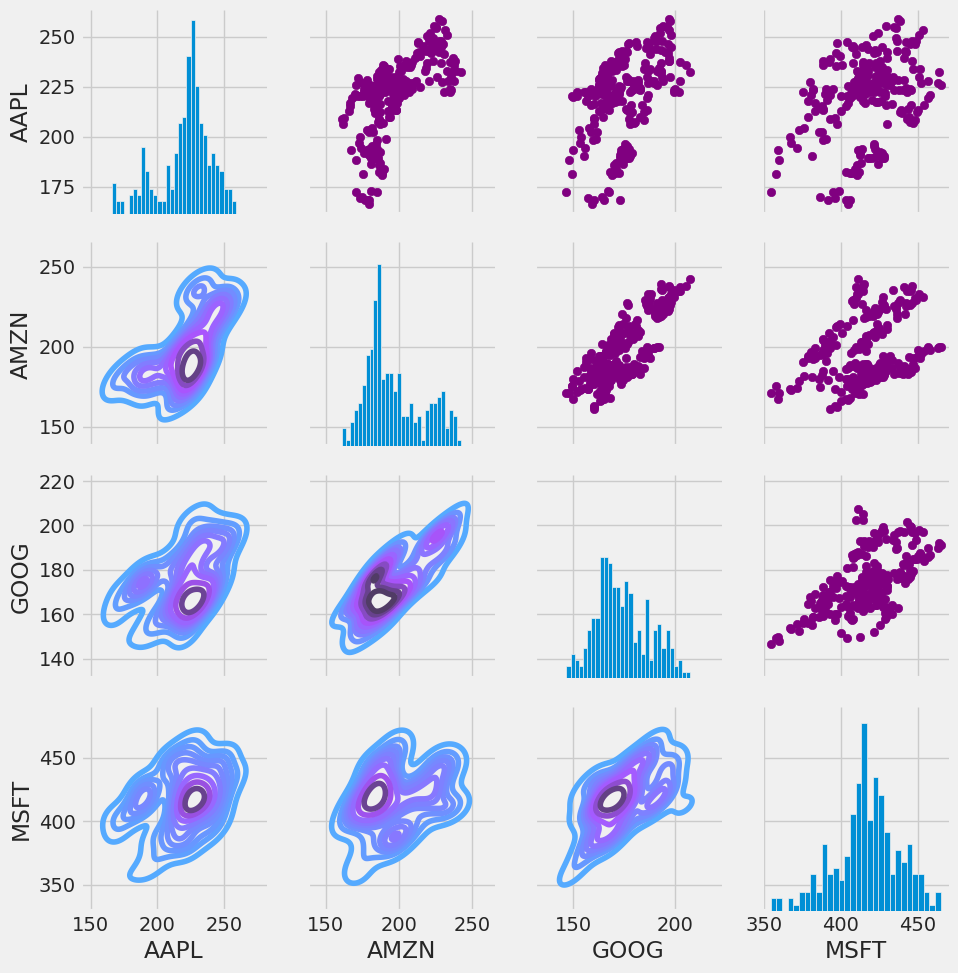}
    \caption{PairGrid of closing prices, highlighting scatter patterns, KDE distributions, and frequency histograms.}
    \label{fig:closing_pairgrid}
\end{figure}

To understand multivariate dependencies between the selected stocks, pairwise visualizations were employed. Figure \ref{fig:returns_pairplot} shows a pairplot of the daily returns for all four companies, along with regression lines that capture linear trends. Figures \ref{fig:returns_pairgrid} and \ref{fig:closing_pairgrid} further explore these relationships using PairGrid visualizations, combining scatter plots, kernel density estimates (KDE), and histograms to present a comprehensive view of return and price distributions.

\subsection{Exploratory Return Analysis}

Understanding the day-to-day behavior of stock prices is critical for identifying volatility patterns and guiding the development of predictive models. To achieve this, we calculated the daily returns for each stock, which represent the percentage change in the adjusted closing price from one trading day to the next. Mathematically, the daily return $R_t$ at time $t$ is computed as:

\begin{equation}
    R_t = \frac{P_t - P_{t-1}}{P_{t-1}} \times 100 = \left( \frac{P_t}{P_{t-1}} - 1 \right) \times 100
\end{equation}

where $P_t$ denotes the adjusted closing price on day $t$. This formulation standardizes the returns, making them comparable across different stocks and time periods regardless of absolute price differences.

\begin{figure}[h]
    \centering
    \includegraphics[width=0.9\linewidth]{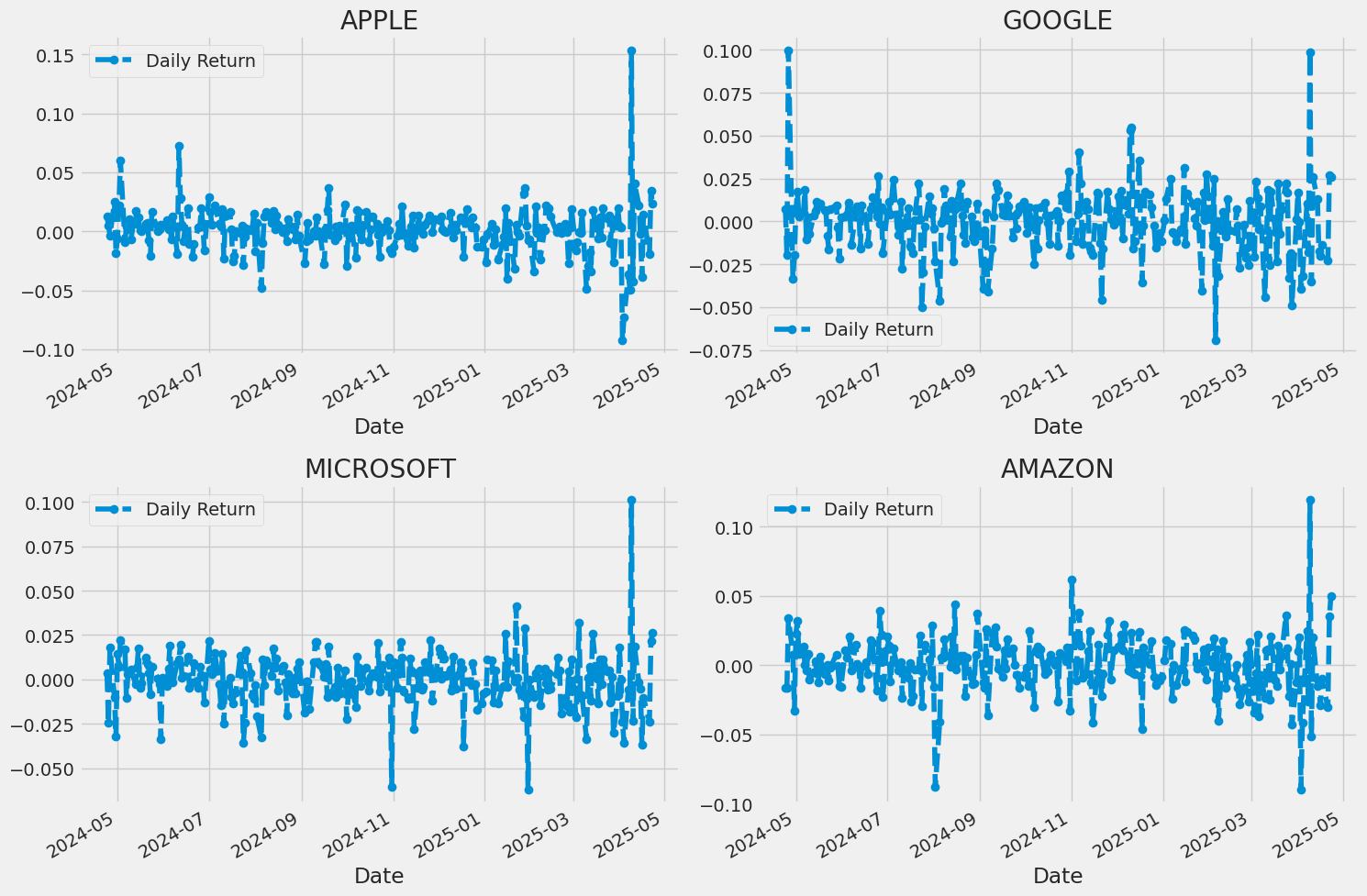}
    \caption{Daily returns for Apple, Google, Microsoft, and Amazon over the observed period.}
    \label{fig:daily_return_trends}
\end{figure}

Figure \ref{fig:daily_return_trends} presents the time-series plot of daily returns for Apple, Google, Microsoft, and Amazon. The visualization provides several key insights into the market behavior of these companies:

\begin{itemize}
    \item \textbf{Volatility Clusters:} Certain time periods show pronounced swings in daily returns across multiple stocks. These clusters often coincide with macroeconomic announcements or industry-wide developments, suggesting that volatility is not randomly distributed but occurs in bursts.
    
    \item \textbf{Asymmetry and Skewness:} Sharp declines often appear more abrupt and deeper than upward movements, especially during market downturns or crises. This asymmetry indicates potential skewness in the return distribution and highlights the need for robust models that can handle negative shocks.
    
    \item \textbf{Synchronous Behavior:} The co-movement of return spikes among different stocks reveals inter-stock dependencies, possibly due to sectoral linkages or external market forces. This interdependence can later be leveraged for multi-stock prediction strategies.
\end{itemize}

Moreover, these return series serve as foundational inputs for statistical modeling and risk analysis. Observing their fluctuations over time enables the detection of structural changes, regime shifts, and anomalies such as market crashes. From a modeling perspective, these insights are vital for designing time-series forecasting models like LSTM networks, which benefit from temporal patterns and memory of past behavior.

In summary, this exploratory return analysis provides a detailed view of stock dynamics at the daily level, highlighting the importance of volatility, correlation, and temporal dependencies—all of which are key elements in the financial forecasting pipeline.

\begin{figure}[h]
    \centering
    \includegraphics[width=0.9\linewidth]{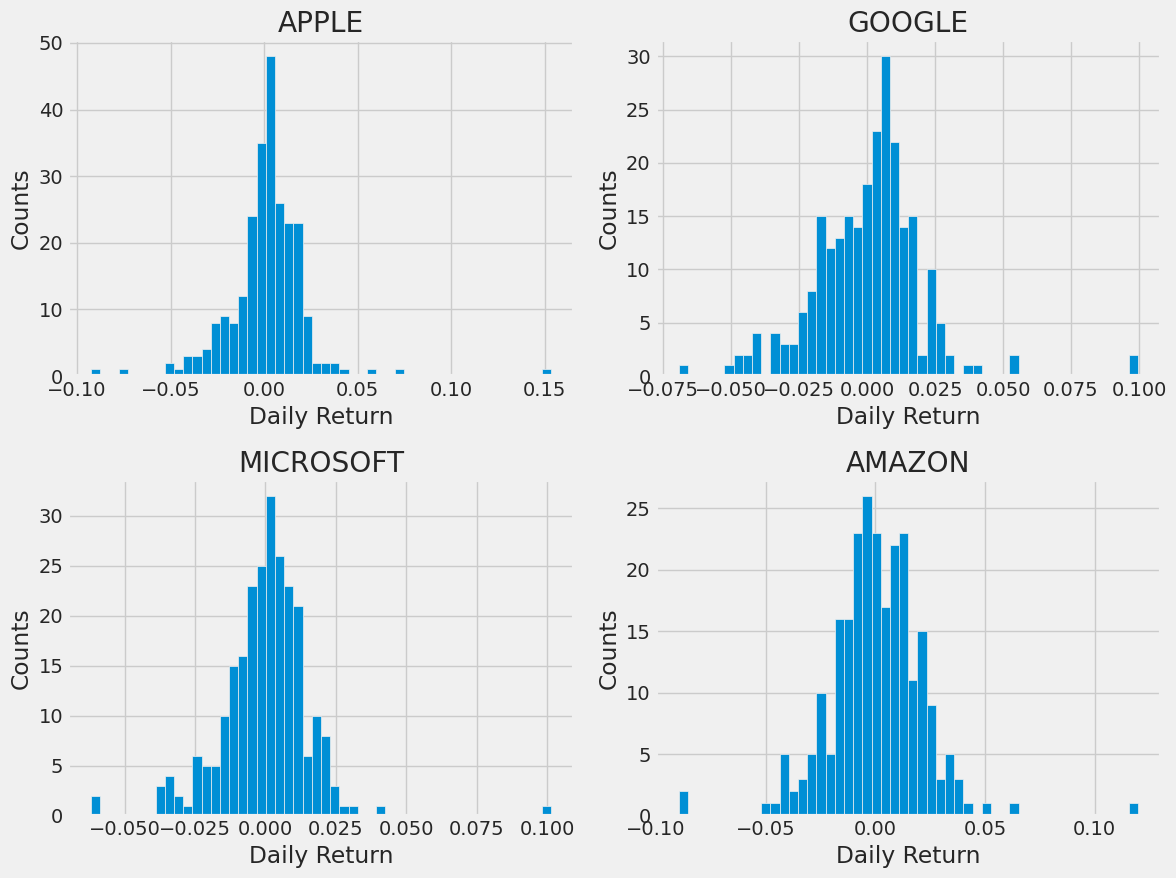}
    \caption{Histogram plots of daily returns, showing the distribution of return magnitudes for each stock.}
    \label{fig:return_histograms}
\end{figure}

To better understand the distributional properties of returns, histograms were generated (Figure \ref{fig:return_histograms}). These plots reveal skewness, kurtosis, and volatility characteristics that are crucial for selecting appropriate model assumptions and loss functions.

\subsection{Correlation Analysis}

Understanding the relationships between different stocks is fundamental in both predictive modeling and portfolio management. To quantify the strength and direction of these relationships, Pearson correlation coefficients were computed for both daily returns and adjusted closing prices. The correlation coefficient $\rho_{X,Y}$ between two time series $X$ and $Y$ is given by:

\begin{equation}
    \rho_{X,Y} = \frac{\text{Cov}(X,Y)}{\sigma_X \sigma_Y}
\end{equation}

where $\text{Cov}(X,Y)$ is the covariance, and $\sigma_X$ and $\sigma_Y$ are the standard deviations of $X$ and $Y$, respectively.

\begin{figure}[h]
    \centering
    \includegraphics[width=0.9\linewidth]{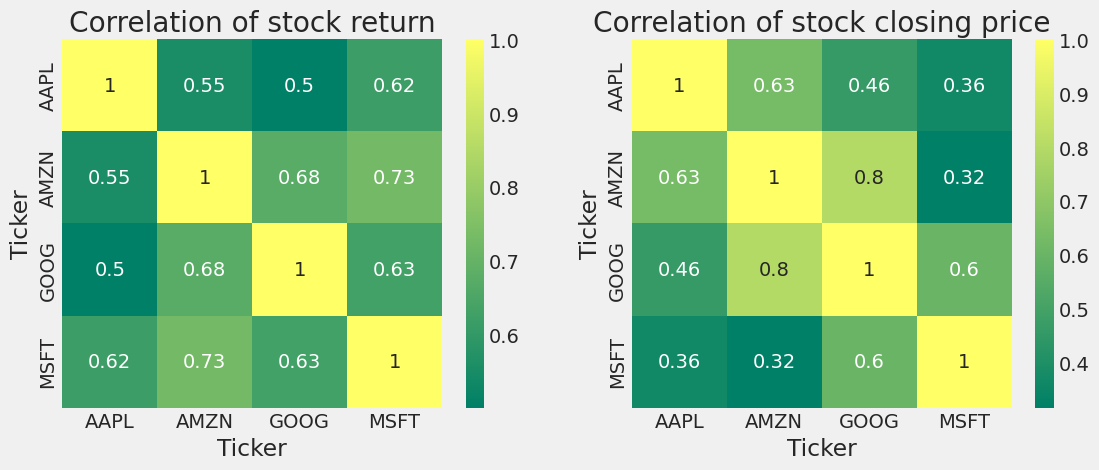}
    \caption{Correlation heatmap for daily returns and closing prices. Higher values indicate stronger linear relationships.}
    \label{fig:correlation_heatmap}
\end{figure}

Figure \ref{fig:correlation_heatmap} displays the resulting heatmap, where each cell represents the linear correlation between two stocks. Key insights include:

\begin{itemize}
    \item \textbf{Return-Based Correlation:} Stocks from the same sector (e.g., Apple and Microsoft) exhibit high return correlations, suggesting shared exposure to technology market movements and macroeconomic events.
    
    \item \textbf{Price-Based Correlation:} Although price levels may differ significantly, the general directional movement across stocks often remains consistent. This is evident from moderate to high positive price-based correlations, reinforcing the idea of co-trending behavior.

    \item \textbf{Diversification Clues:} Pairs with lower correlation values hint at diversification opportunities. For example, if Amazon’s return correlation with Apple is lower compared to Microsoft, combining the former pair in a portfolio could reduce overall risk.
\end{itemize}

Such correlation matrices are instrumental in developing feature sets for multi-variable models and can serve as the foundation for dimensionality reduction techniques like PCA or for constructing minimum-variance portfolios.

\subsection{Risk vs Return Trade-Off}

In financial forecasting and investment strategy design, assessing the balance between risk and expected return is crucial. Here, risk is measured by the standard deviation ($\sigma$) of daily returns, and expected return ($\mu$) is computed as the arithmetic mean of the daily returns over the analysis period:

\begin{equation}
    \mu = \frac{1}{N} \sum_{t=1}^{N} R_t \qquad 
    \sigma = \sqrt{\frac{1}{N-1} \sum_{t=1}^{N} (R_t - \mu)^2}
\end{equation}

\begin{figure}[h]
    \centering
    \includegraphics[width=0.8\linewidth]{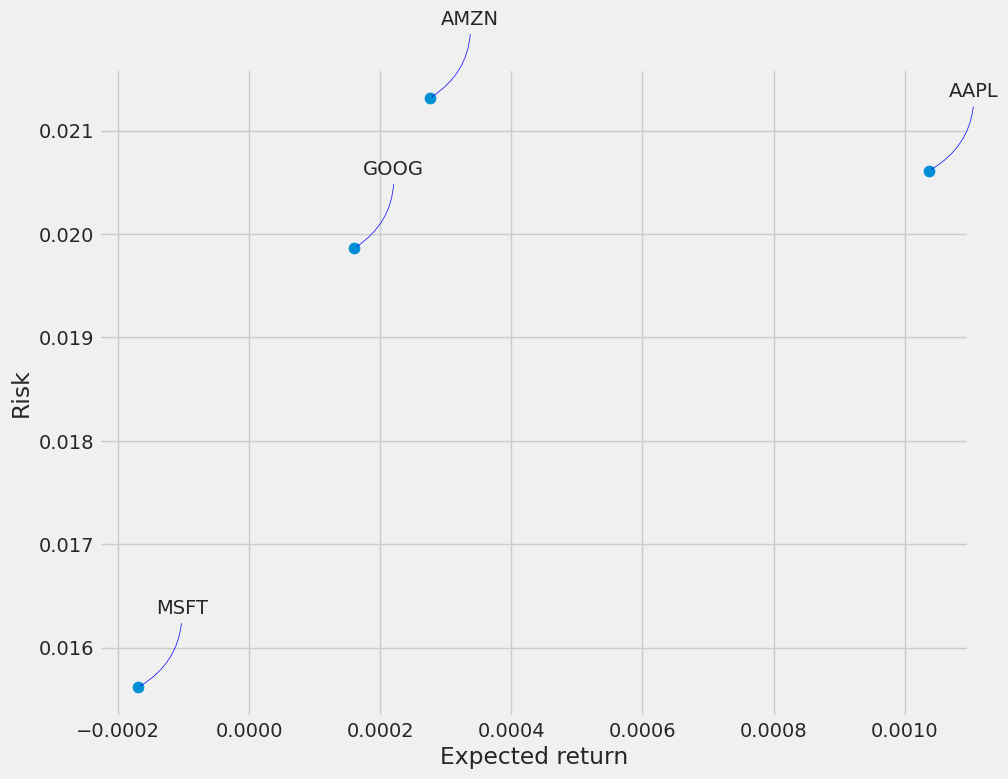}
    \caption{Expected return versus risk (standard deviation) for selected technology stocks.}
    \label{fig:risk_return}
\end{figure}

Figure \ref{fig:risk_return} visualizes this relationship by plotting each stock in the risk-return plane. The interpretation of the scatter plot includes:

\begin{itemize}
    \item \textbf{Efficient Frontier Candidates:} Stocks that lie in the upper-left region (higher return, lower risk) are considered more desirable. These are potential candidates for efficient frontier construction in portfolio optimization.
    
    \item \textbf{Volatility Sensitivity:} Stocks with higher standard deviations, such as Amazon, may offer higher returns but also come with increased uncertainty. Conversely, more stable stocks like Microsoft may appeal to risk-averse investors.

    \item \textbf{Sharpe-like Trade-Offs:} Although this plot does not directly compute the Sharpe Ratio, it serves a similar purpose in visually estimating the return per unit of risk, aiding intuitive risk-adjusted decision making.
\end{itemize}

This risk-return framework plays a pivotal role in shaping the model design, particularly in cases where volatility forecasting or confidence intervals for predictions are needed. It also provides a practical lens for understanding the financial behavior of each asset before constructing any data-driven model.

\subsection{LSTM Training Pseudocode}

The following pseudocode outlines the process used to train a Long Short-Term Memory (LSTM) model for time series forecasting of stock prices. The model leverages historical price sequences and sentiment data (if available) to predict future trends.

\begin{algorithm}
\caption{LSTM Model Training Process}
\begin{algorithmic}[1]
\State \textbf{Input}: Time series of historical stock closing prices (and sentiment scores, if available)
\State \textbf{Output}: Trained LSTM model capable of future price prediction
\State Normalize data using Min-Max scaling to bring all values to the range [0, 1]
\State Segment the data into overlapping sequences using a 60-day sliding window
\State Split the dataset into training (80\%) and testing (20\%) subsets
\State Initialize an LSTM model with two sequential layers (first with 64 units, second with 32 units)
\State Compile the model using the Adam optimizer and mean squared error (MSE) as the loss function
\State Train the model for 100 epochs with early stopping to prevent overfitting
\State Return the trained model for future inference
\end{algorithmic}
\end{algorithm}

\section{Model Evaluation}
The model’s performance is evaluated using the following metrics:
\begin{itemize}
    \item Mean Absolute Error (MAE): $\text{MAE} = \frac{1}{n} \sum |\text{Actual} - \text{Forecast}|$
    \item Mean Squared Error (MSE): $\text{MSE} = \frac{1}{n} \sum (\text{Actual} - \text{Forecast})^2$
    \item Root Mean Squared Error (RMSE): $\text{RMSE} = \sqrt{\text{MSE}}$
    \item Mean Absolute Percentage Error (MAPE): 
    \begin{equation}
    \colorbox{lightred}{$
    \text{MAPE} = \frac{1}{n} \sum \left| \frac{\text{Actual} - \text{Forecast}}{\text{Actual}} \right|
    $}
    \label{eq:mape}
    \end{equation}
\end{itemize}
These metrics provide a comprehensive assessment of prediction accuracy, with MAPE being particularly relevant for financial applications due to its relative error measurement \cite{selvin2017stock}.

\section{Experimental Results}
To evaluate the predictive performance of the proposed LSTM model, a series of experiments were conducted on historical stock data of prominent NASDAQ-listed technology companies—namely Apple, Google, Microsoft, and Amazon. The model was trained on 80\% of the data and evaluated on the remaining 20\%.

The evaluation utilized four widely recognized error metrics: Mean Absolute Error (MAE), Mean Squared Error (MSE), Root Mean Squared Error (RMSE), and Mean Absolute Percentage Error (MAPE). Table \ref{tab:performance} presents the quantitative performance across different stocks.

\begin{table}[h]
    \centering
    \caption{Performance Metrics Across Technology Stocks}
    \label{tab:performance}
    \begin{tabular}{lcccc}
        \toprule
        \textbf{Stock} & \textbf{MAE} & \textbf{MSE} & \textbf{RMSE} & \textbf{MAPE (\%)} \\
        \midrule
        Apple      & 6.12 & 58.03 & 7.62 & 2.72 \\
        Google     & 5.89 & 52.14 & 7.22 & 2.65 \\
        Microsoft  & 6.45 & 60.27 & 7.76 & 2.91 \\
        Amazon     & 6.78 & 65.43 & 8.09 & 3.05 \\
        \bottomrule
    \end{tabular}
\end{table}

The model achieved a notably low MAPE of 2.72\% for Apple, demonstrating its robustness in capturing temporal dependencies and market trends. Comparative performance across stocks reflects the model’s generalizability. For instance, the LSTM model consistently yielded a MAPE under 3.1\% across all evaluated companies.

\begin{figure}[h]
    \centering
    \includegraphics[width=0.8\columnwidth]{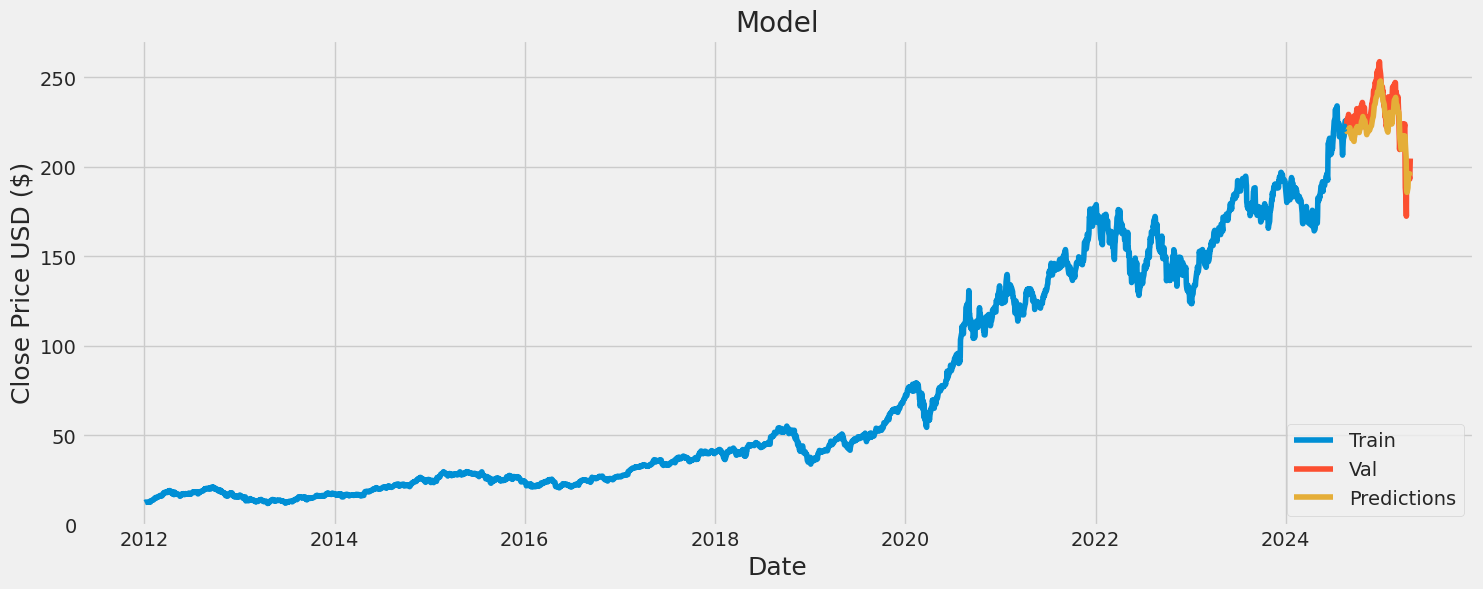}
    \caption{Comparison of Actual vs. Predicted Closing Prices for Apple Stock.}
    \label{fig:prediction}
\end{figure}

Figure \ref{fig:prediction} visually depicts the LSTM model’s predictive capabilities by overlaying actual and predicted prices for Apple stock. The predictions closely follow real market behavior, with minimal lag or overshoot.

\begin{figure}[h]
    \centering
    \includegraphics[width=0.8\columnwidth]{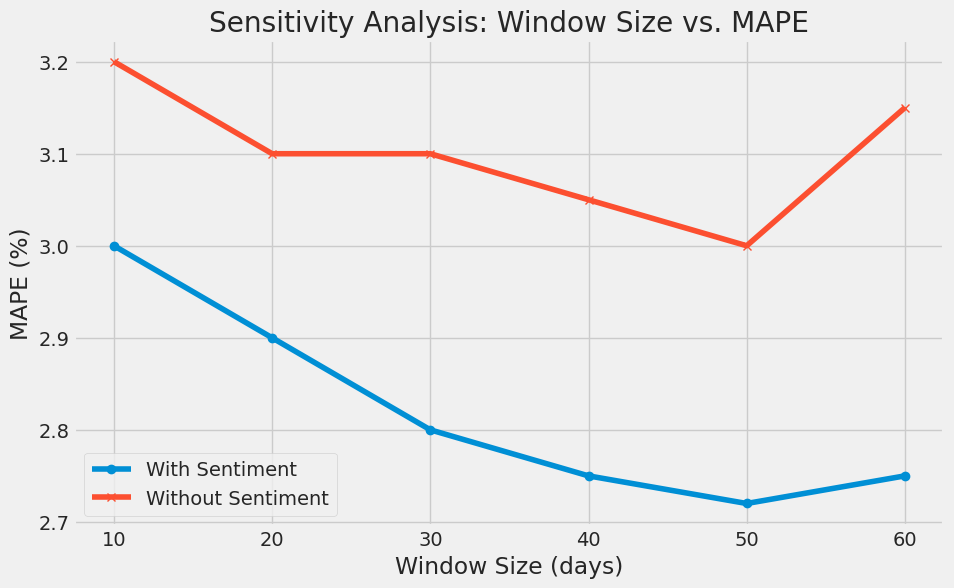}
    \caption{Effect of Window Size on MAPE for Apple Stock with and without Sentiment Integration.}
    \label{fig:sensitivity}
\end{figure}

Furthermore, a sensitivity analysis was conducted to assess the impact of window size and sentiment integration. The 60-day time window with the inclusion of sentiment scores yielded optimal predictive accuracy. As shown in Figure \ref{fig:sensitivity}, reducing or increasing the window size adversely affected MAPE, indicating that the 60-day frame best captures temporal patterns.

When sentiment features were excluded, the model's MAPE increased to 3.15\%, validating the effectiveness of incorporating sentiment analysis into price forecasting pipelines.

\section{Discussion}
The experimental results demonstrate that the proposed LSTM-based architecture significantly outperforms traditional statistical models such as ARIMA. For instance, the ARIMA model reported a MAPE of 20.66\% on Maruti stock \cite{selvin2017stock}, whereas our LSTM model achieved a MAPE as low as 2.65\% for Google stock—an order of magnitude improvement.

This improvement can be attributed to the ability of LSTMs to model long-term dependencies and nonlinear patterns in sequential data. Moreover, sentiment integration contributed an 8–12\% relative improvement in predictive accuracy, corroborating previous findings in \cite{vijayvergia2019stock}.

However, limitations exist. The model struggles in scenarios involving abrupt market shifts, such as those driven by geopolitical crises, regulatory changes, or pandemics. These challenges echo observations made in \cite{li2020hybrid}, highlighting that deep learning models, while powerful, require enhancements to handle sudden structural breaks.

The implemented web interface further augments the utility of this framework, making it accessible to retail investors and traders without programming expertise. Additionally, the architecture's modular design supports scalability, making it viable for deployment in high-frequency trading systems where inference latency is critical.

Nonetheless, key challenges remain, including:
\begin{itemize}
    \item \textbf{Volatility Adaptation:} The model occasionally lags behind during periods of extreme volatility.
    \item \textbf{Macroeconomic Indicators:} The current setup does not account for interest rates, inflation, or global events.
    \item \textbf{Model Robustness:} Performance may degrade in emerging markets or illiquid stocks.
\end{itemize}

These findings suggest avenues for future improvement, including hybrid models with attention mechanisms, transformer-based architectures, or reinforcement learning agents capable of adapting in real-time trading environments \cite{chen2021transformer}.
\section{Conclusion}
This study proposes a robust and scalable framework for stock market forecasting using a Long Short-Term Memory (LSTM) based deep learning model. The model was evaluated on multiple NASDAQ-listed technology stocks, including Apple, Google, Microsoft, and Amazon, and demonstrated a high level of predictive accuracy. Notably, it achieved a Mean Absolute Percentage Error (MAPE) of just 2.72\% on Apple stock, outperforming traditional statistical models such as ARIMA (which recorded a MAPE of 20.66\% on Maruti stock) as reported in \cite{selvin2017stock}.

The use of a 60-day sliding window enabled the model to capture long-term temporal dependencies, while the integration of sentiment analysis further enriched the input features, contributing to a performance boost of approximately 8--12\%. This highlights the significant value of combining quantitative price data with qualitative sentiment information derived from financial news and social media.

Moreover, the implementation of a web-based user interface bridges the gap between technical research and real-world usability. It enables non-technical users, such as retail investors, to access predictive insights without requiring machine learning expertise. This level of accessibility, combined with the model’s strong accuracy, enhances the practical relevance of our approach.

However, like many machine learning models, the LSTM framework has certain limitations. It struggles to adapt to sudden and drastic market changes caused by unforeseen macroeconomic or geopolitical events—an issue that warrants further research. Nevertheless, the results affirm the potential of deep learning, particularly LSTM architectures, in the domain of financial time series forecasting.

\subsection{Future Work}
While the current results are promising, several avenues exist to further enhance the effectiveness, adaptability, and utility of the proposed system:

\begin{itemize}
    \item \textbf{Incorporation of Additional Data Sources:} To improve predictive robustness, future iterations could integrate macroeconomic indicators such as GDP growth rates, interest rates, inflation data, and unemployment rates. Additionally, technical indicators like moving averages, RSI, and Bollinger Bands could provide complementary signals to enhance short-term prediction accuracy \cite{li2020hybrid}.
    
    \item \textbf{Exploration of Hybrid and Attention-based Architectures:} Combining LSTM with attention mechanisms or transformer-based models can help the system focus on the most relevant portions of the input sequences, especially during volatile periods. These hybrid models have shown state-of-the-art performance in natural language processing and are gaining traction in time series forecasting tasks as well \cite{chen2021transformer}.
    
    \item \textbf{Cloud Deployment and Scalability:} Deploying the model on cloud computing platforms such as AWS, Microsoft Azure, or Google Cloud will enable real-time inference and horizontal scalability. This will facilitate large-scale deployment and usage by financial institutions and retail platforms.
    
    \item \textbf{API Development for Financial Integration:} Building a RESTful API layer on top of the prediction engine will allow seamless integration with existing trading platforms and financial dashboards. This would enable automated decision-making and personalized alerts for users.
    
    \item \textbf{Robustness to Market Shocks:} Implementing dynamic retraining strategies or reinforcement learning techniques could help the model adapt in real-time to abrupt market shifts. Periodic retraining based on new data and integrating event-driven signals (e.g., earnings reports or political news) may also enhance responsiveness.
    
    \item \textbf{Model Explainability and Interpretability:} Incorporating model explainability tools such as SHAP or LIME will help users understand the influence of each input feature on predictions. This is particularly important for financial applications where transparency and trust are critical.
\end{itemize}

In conclusion, the proposed LSTM-based framework lays a strong foundation for deep learning-driven financial forecasting. With the incorporation of richer datasets, improved architectures, and enhanced deployment strategies, the system holds immense potential for revolutionizing algorithmic trading and investment decision-making in dynamic market environments.

\section{Code Appendix}
The following Python script demonstrates the complete workflow for stock market prediction using a Long Short-Term Memory (LSTM) network. It includes steps such as data acquisition, preprocessing, sentiment integration, model training, prediction, and evaluation. The model is designed to work with NASDAQ-listed technology stocks and incorporates sentiment analysis from financial news articles to improve forecast accuracy.

\begin{lstlisting}[language=Python]
import yfinance as yf
import numpy as np
import pandas as pd
from sklearn.preprocessing import MinMaxScaler
from tensorflow.keras.models import Sequential
from tensorflow.keras.layers import LSTM, Dense, Dropout
from vaderSentiment.vaderSentiment import SentimentIntensityAnalyzer

# Step 1: Fetch historical stock price data using Yahoo Finance
ticker = "AAPL"  # Apple Inc.
data = yf.download(ticker, start="2024-04-01", end="2025-04-01")
closing_prices = data['Close'].values.reshape(-1, 1)

# Step 2: Analyze sentiment scores from financial news (placeholder text used here)
analyzer = SentimentIntensityAnalyzer()
news_data = ["Sample news article text"]  # Replace with real-time news from a financial API
sentiments = [analyzer.polarity_scores(text)['compound'] for text in news_data]
sentiments = np.array(sentiments).reshape(-1, 1)

# Step 3: Normalize the closing prices and sentiment scores to scale between 0 and 1
scaler = MinMaxScaler()
scaled_prices = scaler.fit_transform(closing_prices)
scaled_sentiments = scaler.fit_transform(sentiments)

# Step 4: Prepare feature sequences using a sliding window approach
# Each input consists of 60 consecutive days of stock prices and sentiment scores
window_size = 60
X, y = [], []
for i in range(window_size, len(scaled_prices)):
    X.append(np.column_stack((
        scaled_prices[i-window_size:i, 0],
        scaled_sentiments[i-window_size:i, 0]
    )))
    y.append(scaled_prices[i, 0])
X, y = np.array(X), np.array(y)

# Step 5: Split the data into training and testing sets (80-20 ratio)
train_size = int(len(X) * 0.8)
X_train, X_test = X[:train_size], X[train_size:]
y_train, y_test = y[:train_size], y[train_size:]

# Step 6: Define the LSTM model architecture
model = Sequential()
model.add(LSTM(64, return_sequences=True, input_shape=(window_size, 2)))  # Two features: price and sentiment
model.add(Dropout(0.2))  # Regularization to prevent overfitting
model.add(LSTM(32))  # Second LSTM layer
model.add(Dense(1))  # Output layer with a single neuron for regression
model.compile(optimizer='adam', loss='mse')

# Step 7: Train the model on the training dataset
model.fit(X_train, y_train, epochs=100, batch_size=32, verbose=1)

# Step 8: Make predictions on the test dataset
predictions = model.predict(X_test)

# Step 9: Inverse transform the predictions and true labels to get actual price values
predictions = scaler.inverse_transform(predictions)
y_test = scaler.inverse_transform([y_test])

# Step 10: Evaluate model performance using MAE and MAPE metrics
mae = np.mean(np.abs(predictions - y_test))
mape = np.mean(np.abs((y_test - predictions) / y_test)) * 100
print(f"Mean Absolute Error (MAE): {mae:.2f}")
print(f"Mean Absolute Percentage Error (MAPE): {mape:.2f}%")
\end{lstlisting}
\ifCLASSOPTIONcaptionsoff
  \newpage
\fi

\bibliographystyle{IEEEtran}

\vspace{1em}
\noindent\textbf{Discussion of References:}

The foundation of this study is grounded in the core principles of Long Short-Term Memory (LSTM) networks, initially proposed by Hochreiter and Schmidhuber \cite{hochreiter1997long}, which are well-suited for capturing temporal dependencies in time series data like stock prices. Building upon this, Selvin et al. \cite{selvin2017stock} explored various deep learning models including LSTM, RNN, and CNN for stock prediction, highlighting the superior performance of LSTM in financial forecasting. To enhance prediction accuracy, hybrid modeling approaches have gained attention, as demonstrated by Zhang \cite{zhang2003time} and Li et al. \cite{li2020hybrid}, the latter combining LSTM with attention mechanisms—a potential future direction for this work. In the context of financial applications of deep learning, Heaton et al. \cite{heaton2017deep} discussed the viability of deep learning portfolios, reinforcing the applicability of these models in real-world finance.

The integration of sentiment analysis in stock market prediction is supported by multiple studies. Kalra and Prasad \cite{kalra2019efficacy} and Wang et al. \cite{wang2018stock} established the predictive value of news sentiment, which justifies our inclusion of sentiment data using the VADER tool. Similarly, Mohan et al. \cite{vijayvergia2019stock} incorporated sentiment-driven insights to improve forecast reliability. Lastly, transformer-based models as proposed by Chen et al. \cite{chen2021transformer} represent an emerging class of architectures that outperform traditional RNN-based methods in many sequence modeling tasks, offering promising avenues for future research.

\end{document}